# Isabelle: The Next 700 Theorem Provers

Lawrence C Paulson
Computer Laboratory
University of Cambridge
Pembroke Street
Cambridge CB2 3QG
United Kingdom

# Contents

# 1 Introduction

Isabelle is a *generic* theorem prover, designed for interactive reasoning in a variety of formal theories. At present it provides useful proof procedures for Constructive Type Theory (Per Martin-Löf, 1984), various first-order logics, Zermelo-Fraenkel set theory, and higher-order logic. This survey of Isabelle serves as an introduction to the rather formidable literature.

Generic theorem proving may seem a foolish enterprise. When theorem proving for a fixed logic is already so difficult, why complicate matters by letting the logic vary? The answer is that many of the difficulties have to do with logic in general. These can be dealt with once and for all, leaving the peculiarities of individual logics to be dealt with separately. The resulting system may be far less powerful than a specialized theorem prover, but will be more flexible. Many automatic systems force the user to work with a fixed search strategy as well as a fixed logic.

Isabelle embraces logics from a fairly broad family. Some members of the family are constructive, others classical. They include first-order and higher-order logics. Some are based on sets, others on types and functions, others on domains. The family is evolving: new members are born, others develop and mature, and some disappear. As it is applied to a particular problem, a logic becomes more and more specialized, and Isabelle accommodates this. Distant cousins, however, such as relevance and dynamic logics, may fall outside Isabelle's scope. It may not be possible to accommodate everyone.

We shall start with a thorough history of Isabelle, beginning with its origins in LCF. This is a tale of errors, not of a grand design. There follows an account of how logics are represented, illustrated using classical logic. The approach is compared with the Edinburgh Logical Framework. Several of the Isabelle object-logics are then presented, and finally the conclusions survey work in progress.

# 2 The LCF approach to theorem proving

First we should recall the history of LCF. Around 1972, Robin Milner developed a proof checker for Scott's Logic for Computable Functions. His results led him to seek a compromise between fully automatic theorem proving, which seemed impossibly difficult, and single step proof checking, which seemed impossibly tedious. Around 1977, his group developed a *programmable* proof checker, Edinburgh LCF. Inference rules were expressed as functions in a programmable meta-language (called ML). By writing programs in ML, users could automate the proof process as much as desired.

LCF also permits a proof to be constructed backwards from a *goal*, the conjecture to be proved. An LCF *tactic* is a function that reduces a goal to zero or more subgoals. Once all the subgoals have been solved, LCF (using complex bookkeeping) constructs the corresponding forwards proof, yielding the desired theorem. *Tacticals* combine tactics in various ways. Tactics and tacticals constitute a powerful control language; they can describe search methods such as 'repeatedly apply Rule $X$ then repeatedly apply either Rule $Y$ or Rule $Z$'.

Edinburgh LCF proved a great success. Many similar systems were developed: including Cambridge LCF (Paulson, 1987); Nuprl for Constructive Type Theory (Constable et al., 1986); and HOL for higher-order logic (Gordon, 1988).

In 1985 there were already two LCF-style theorem provers for Constructive Type Theory. In writing a third, I had specific aims. One was to incorporate Sokołowski's (1987) technique for solving unknowns in goals by unification; this involved a complicated treatment of variable assignments in tactics. Another aim was to write the entire system in Standard ML — which had grown out of LCF's meta-language — to show it was a practical alternative to Lisp. (For an introduction, see Wikström (1987).) Another aim was to experiment with de Bruijn's (1972) treatment of bound variables. The data structure was Martin-Löf's theory of expressions (similar to the typed λ-calculus), which seemed able to represent most forms of logical syntax.

Making an LCF-style system was time-consuming and uncertain. Every year or so, LCF users found some serious bug. (Most were due to bound variable clashes in strange situations.) There was widespread concern that computer scientists could not implement logics as fast as logicians could conceive them. The ideal solution seemed to be a generic theorem prover for a broad class of logics. Realistically, the most one could aim to provide was a standard library of syntactic operations. This would spare implementors from coding tricky substitution functions, but they would still have to write a function and a tactic for each inference rule.

## 3 Proof construction by unification

Unification has long been important in theorem proving. It seems essential for Constructive Type Theory, where we typically have to prove $?a \in A$, where $A$ represents a proposition and $?a$ is an unknown. At the end of the proof, the meta-variable $?a$ must be replaced by some proof object. A series of unifications could construct this object step by step.

But there is a snag. Although standard unification can easily be modified to cope with bound variables, the Type Theory proof process builds λ-abstractions, and so unification must instantiate function variables.

Unaware of this snag, I foolishly wrote a mountain of code, and it did not work. What was required was *higher-order unification*: a special unification procedure for the typed λ-calculus, with its bound variables and conversion rules. Gérard Huet (1975) had already designed such a procedure, and it had been successfully applied by Andrews et al. (1984). A quick implementation showed that higher-order unification could perform the syntactic operations of rules and tactics. An inference rule could be represented literally: as a list of premises and a conclusion. The meta-variables in rules would be schematic variables for higher-order unification. Proof construction would work by combining rules, unifying the conclusion of one rule with a premise of another. This is the basic idea of Isabelle.

Higher-order unification has appalling theoretical properties: it is undecidable and the set of solutions may be infinite. In practice, Huet's procedure works reliably. Perhaps Isabelle uses only easy cases.

Quantifiers cause the main difficulties. Martin-Löf (1984) showed how complicated variable binding and substitution rules can be uniformly presented in (essentially) the typed λ-calculus. But what about *eigenvariables*, the special variables in quantifier rules? Consider the eigenvariable $x$ in the rule

$$\frac{P[x]}{\forall x.P[x]}$$

The eigenvariable proviso, '$x$ not free in the hypotheses', can be enforced in numerous ways. One idea was to replace $x$ by a Skolem term (Paulson, 1986). A related idea, based on directed acyclic graphs, worked much better and led to the first usable version of Isabelle. Now Isabelle uses yet another approach, described in Section 6.3.

## 4 Early experiments

At first Isabelle was horribly slow. In the worst example, solving trivial subgoals took minutes of computer time. There was plainly much re-traversal of redundant subexpressions. Complicated 'structure sharing' techniques were tried without success. The cure turned out to be simple: eliminate Skolem terms, the cause of the redundancy. Another case of poor performance was finally traced to a subtle heuristic defect in the unification search procedure; the cure involved a single line of code.

Though efficiency should not become an obsession, the program has to run fast enough to be usable. The issues are the same as in conventional programming, and so are the techniques. ML's execution profiler reported that the sharing mechanism, meant to boost efficiency, was consuming most of the run time. The replacement of structure sharing by copying made Isabelle simpler and faster. Complex algorithms are often the problem, not the solution.

By 1986, Isabelle could accomplish automated proof using tactics and tacticals more powerful than LCF's. There was no need to write a tactic for each rule since a standard tactic applied any given rules to a goal. Goals could contain variables, even function variables, to be solved by unification. If all this was not enough, the user always had recourse to ML.

The 1986 version of Isabelle is called Isabelle-86 to distinguish it from the present Isabelle, which uses a different representation of rules (described in the next section). The object-logics of Isabelle-86 can now be discussed.

For Constructive Type Theory many useful proof techniques are implemented, such as type inference and term rewriting. Rewriting can be easily expressed in pure PROLOG — it is just a form of equality reasoning. The same idea works for Constructive Type Theory, where rewriting is used for many complicated proofs about arithmetic.

First-order logic is implemented in several versions. Thanks to a devious encoding of lists as nested function applications, higher-order unification can achieve the effect of associative unification (Huet and Lang, 1978). Thus we may formulate the sequent calculus using sequence variables. For example, a *basic sequent* is one that has a formula in common on the left and right sides; every basic sequent is obviously valid. A scheme for basic sequents can be directly represented in Isabelle:

$$?\Gamma, ?P, ?\Gamma' \vdash ?\Delta, ?P, ?\Delta'$$

Again, a question mark indicates a meta-variable. Note that a meta-variable in a goal represents an unknown, while meta-variables in a rule express its schematic structure.

Here $?\Gamma$, $?\Delta$, $?\Gamma'$, and $?\Delta'$ denote sequences while $?P$ denotes a formula. Unifying this scheme with the sequent $Q(?a), Q(f(?b)) \vdash Q(f(1))$ produces two unifiers: one with $?a = f(1)$ and one with $?b = 1$. Each unifier gives a valid instance of the sequent. A naive procedure of repeated rule application can prove moderately hard examples automatically: about 35 of the problems in Pelletier (1986).

Zermelo-Fraenkel set theory has been implemented over first-order logic. Set theory is defined through first-order axioms. These are used to derive a family of rules, forming a sequent calculus for set theory. The resulting proof procedure can prove many interesting facts.

## 5 Extending the rule calculus

If the naive rule calculus could cope with the range of mathematical reasoning from Constructive Type Theory to Zermelo-Fraenkel set theory, why consider more complex logical frameworks? One problem was that Isabelle-86 did not support natural deduction. Each logic had to be formulated as a sequent calculus with explicit rules for assumptions. In practice, though, this seemed a minor problem.

The real puzzle was how best to support the use of derived rules. Derived rules permitted working at an abstract level and produced shorter proofs. To prove a *theorem* Isabelle-86 worked well; to derive a *rule* required reducing the conclusion to a given set of premises, which was as simple as threading a needle with a piece of wet spaghetti. The solution seemed to involve taking the desired premises as meta-level assumptions and using them in the proof. A further complication: Constructive Type Theory required taking rules as assumptions.

So the Isabelle-86 rule calculus needed a radical extension. In Schroeder-Heister's (1984) 'rules of higher level', a rule could discharge assumptions, and each assumption could itself be a schematic rule. But there were extremely complicated discharge functions and variable conditions. Unpublished work by Martin-Löf developed similar concepts using 'propositions as types'; people at Edinburgh formalized these ideas as the Logical Framework.

It is not difficult to capture the same ideas in a framework better suited to Isabelle. To the typed $\lambda$-calculus add implication and a typed universal quantifier with their usual meanings. We obtain a meta-level inference system for proof construction, replacing the old calculus of object-rules. Implication expresses logical entailment (including assumption discharge) while the quantifier expresses generality (including eigenvariable conditions). When rules are combined, assumptions and eigenvariables accumulate properly. The new system is much better with derived rules, especially under natural deduction; these two concepts appear to be closely related.

Isabelle's new meta-logic is the fragment of higher-order logic (Andrews, 1986) with implication ($\Longrightarrow$), the universal quantifier ($\bigwedge$), and equality ($\equiv$). Non-standard symbols are used to leave the standard symbols free for the object-logics. The type of propositions (Church's type $o$) is called *prop*. None of the existing formalizations of logics require quantification over *prop*, only quantification over object truth-values. There are philosophical and technical reasons for prohibiting abstractions involving *prop*, staying within predicative simple type theory. For a detailed discussion, see Paulson (1989).

The rules are shown to allow comparison with the Edinburgh Logical Framework. Bear in mind this is not a new system but part of an old one (Church, 1940), expressed in natural deduction style.

| introduction | elimination |
| --- | --- |
| $\dfrac{\begin{matrix}[\phi]\\ \psi\end{matrix}}{\phi \Longrightarrow \psi}$ | $\dfrac{\phi \Longrightarrow \psi \qquad \phi}{\psi}$ |
| $\dfrac{\phi[x]}{\bigwedge x.\phi[x]}\,*$ | $\dfrac{\bigwedge x.\phi[x]}{\phi[t]}$ |

* Proviso: the variable $x$ must not be free in the assumptions.

Here $\phi[x]$ indicates that the formula $\phi$ may have free occurrences of $x$. In this context $\phi[t]$ means the substitution of $t$ for $x$ in $\phi$, subject to the usual conditions. The $\bigwedge$ introduction rule has an eigenvariable condition. In $\bigwedge$ elimination, term $t$ and variable $x$ must have the same type.

Isabelle-86 permits constants to be defined, but the new meta-logic expresses definitions using meta-level equality ($\equiv$). The axiom $C \equiv t$ defines the constant $C$ to equal the term $t$. By the usual equality properties, we can replace $C$ by $t$ (or the reverse) in a proof. The meta-logic includes the equality rules of the typed $\lambda$-calculus, such as substitution and $\alpha$, $\beta$, and $\eta$-conversion.

A logic is represented by introducing new types, constants, and axioms. The representation can be viewed syntactically or semantically. In the syntactic viewpoint, certain strings of symbols in the object-logic are mapped to strings of symbols in the meta-logic. This viewpoint is convenient for proving formal properties of the representation, such as soundness.

The semantic viewpoint observes that the meta-logic (higher-order logic) has simple models. Each type denotes a set, each typed term denotes an element of the corresponding set, and the logical constants have their usual meanings. If the object-logic also has a semantics, then the representation in higher-order logic ought to preserve this semantics. This point of view is helpful for informal mathematical reasoning. If the object-logic uses a syntactic device that is awkward for the meta-logic, then it can be formalized differently. Often such changes of formalization are 'obviously' reasonable, though the formal equivalence in the model theory could be difficult to prove.

## 6 A formalization of first-order logic

As an example of the meta-logic, let us represent classical first-order logic. The following is a standard natural deduction system (Prawitz, 1965), where $\bot$ means falsity and $\neg P$ abbreviates $P \supset \bot$.

| introduction | elimination |
| --- | --- |
| $\dfrac{P \quad Q}{P \wedge Q}$ | $\dfrac{P \wedge Q}{P} \qquad \dfrac{P \wedge Q}{Q}$ |
| $\dfrac{P}{P \vee Q} \qquad \dfrac{Q}{P \vee Q}$ | $\dfrac{P \vee Q \quad \begin{matrix}[P] \\ R\end{matrix} \quad \begin{matrix}[Q] \\ R\end{matrix}}{R}$ |
| $\dfrac{\begin{matrix}[P] \\ Q\end{matrix}}{P \supset Q}$ | $\dfrac{P \supset Q \quad P}{Q}$ |
| | $\dfrac{\begin{matrix}[P \supset \bot] \\ \bot\end{matrix}}{P}$ |
| $\dfrac{P[x]}{\forall x.P[x]}*$ | $\dfrac{\forall x.P[x]}{P[t]}$ |
| $\dfrac{P[t]}{\exists x.P[x]}$ | $\dfrac{\exists x.P[x] \quad \begin{matrix}[P[x]] \\ Q\end{matrix}}{Q}\dagger$ |

* Proviso: the variable $x$ must not be free in the assumptions.
† Proviso: the variable $x$ must not be free in $Q$ or in any assumption save $P[x]$

## 6.1 The treatment of syntax

First-order logic deals with terms and formulae. To represent these in the typed $\lambda$-calculus, we introduce types *term* and *form* and constant symbols for the connectives and quantifiers.

$$\begin{aligned} \bot &\in form \\ \wedge, \vee, \supset &\in form \rightarrow (form \rightarrow form) \\ \forall, \exists &\in (term \rightarrow form) \rightarrow form \end{aligned}$$

Here $\bot \in form$ means '$\bot$ has type *form*'. We will write $\wedge$, $\vee$, $\supset$ as infixes; note that they are curried functions of two arguments. The types of $\forall$ and $\exists$ are explained below.

In the syntactic viewpoint, every meta-level term of type *term* encodes some first-order term, and every meta-level term of type *form* encodes some formula. The logical constants construct formulae: if $P$ and $Q$ are codings of formulae then so is $P \wedge Q$. Furthermore, $\lambda$-abstraction encodes the use of bound variables in quantifiers: the encoding of $\forall x.P[x]$ is $\forall(\lambda x.P[x])$.

To understand the representation of the rules we shall need the semantic viewpoint. Type *form* denotes the set of truth values, $\{\bot, \top\}$. The connectives $\wedge$, $\vee$, $\supset$ denote Boolean functions defined by truth tables. Type *term* denotes a non-empty set of individuals: in first-order logic, the universe is assumed to be non-empty. As it happens, all types denote non-empty sets in Isabelle's version of higher-order logic.

The quantifiers $\forall$ and $\exists$ are infinitary versions of $\wedge$ and $\vee$. Suppose $P[x]$ has type *form* where $x$ has type *term*. If $P[x]$ is true (equals $\top$) for all $x$, then $\lambda x.P[x]$ is a function on individuals that is constantly equal to $\top$. In this case $\forall(\lambda x.P[x])$ equals $\top$. More generally, $\forall(F)$ equals $\top$ if and only if $F(x)$ equals $\top$ for all $x$. The existential quantifier can similarly be interpreted using an infinite truth table to define $\exists(F)$.

Although object-level truth (type *form*) should be distinguished from meta-level truth (type *prop*), they are related. The meta-level only needs to know whether a given value in *form* is true. We could introduce the constant $\top$ and express '$P$ is true' by $P \equiv \top$, but it is more general to introduce the meta-predicate that holds of all truths:

$$\mathrm{true} \quad : \quad \mathit{form} \to \mathit{prop}$$

Then '$P$ is true' becomes $\mathrm{true}(P)$, usually abbreviated as $[\![P]\!]$. Now for each object-level rule we introduce an axiom.

## 6.2 The formalization of the rules

The conjunction rules are simple, and the corresponding axioms can be read by the obvious syntactic translation. The semantic reading is also simple: 'if $P \wedge Q$ is true then $P$ is true', and so forth. Note that $\Longrightarrow$ associates to the right; the parentheses just below could be omitted.

$$\bigwedge PQ \,.\, [\![P]\!] \Longrightarrow ([\![Q]\!] \Longrightarrow [\![P \wedge Q]\!])$$

$$\bigwedge PQ \,.\, [\![P \wedge Q]\!] \Longrightarrow [\![P]\!] \qquad\qquad \bigwedge PQ \,.\, [\![P \wedge Q]\!] \Longrightarrow [\![Q]\!]$$

For disjunction, the introduction rules are simple but the elimination rule discharges assumptions. This is discussed below.

$$\bigwedge PQ \,.\, [\![P]\!] \Longrightarrow [\![P \vee Q]\!] \qquad\qquad \bigwedge PQ \,.\, [\![Q]\!] \Longrightarrow [\![P \vee Q]\!]$$

$$\bigwedge PQR \,.\, [\![P \vee Q]\!] \Longrightarrow ([\![P]\!] \Longrightarrow [\![R]\!]) \Longrightarrow ([\![Q]\!] \Longrightarrow [\![R]\!]) \Longrightarrow [\![R]\!]$$

Now consider implication. The elimination rule is simple, while the introduction rule discharges an assumption. The rationale for this rule is that if from assuming $P$ it follows that $Q$ is true, then $P \supset Q$ must be true. The axiom simply formalizes this in the meta-logic, rendering both 'it follows that' and 'if-then' as meta-implications. Syntactically, both rule application and assumption discharge are translated to $\Longrightarrow$.

$$\bigwedge PQ \,.\, ([\![P]\!] \Longrightarrow [\![Q]\!]) \Longrightarrow [\![P \supset Q]\!]$$

$$\bigwedge PQ \,.\, [\![P \supset Q]\!] \Longrightarrow ([\![P]\!] \Longrightarrow [\![Q]\!])$$

Together these axioms define $[\![P \supset Q]\!]$ as equivalent to $[\![P]\!] \Longrightarrow [\![Q]\!]$.

If the contradiction rule is omitted, we have minimal logic. For intuitionistic logic we could formalize a weak contradiction rule:

$$\bigwedge P \,.\, [\![\bot]\!] \Longrightarrow [\![P]\!]$$

The classical rule discharges the assumption $\neg P$:

$$\bigwedge P \,.\, ([\![P \supset \bot]\!] \Longrightarrow [\![\bot]\!]) \Longrightarrow [\![P]\!]$$

Using Isabelle we can define minimal logic and then define intuitionistic and classical logic as extensions of it. We cannot share theorem provers, however: most classical methods are invalid for intuitionistic logic.

### 6.3 Quantifier issues

Quantifier rules are traditionally expressed using a notation for substitution, as in $P[t]$. When the rules are formalized in Isabelle, the scope of the quantifier becomes a function variable, say $F$. Substitution is obtained by function application: $F(t)$. Syntactically, if $F$ is $\lambda x\,.\,P[x]$ then $F(t)$ is $(\lambda x\,.\,P[x])(t)$, which is $P[t]$ by $\beta$-conversion. Semantically, $F$ is a function from individuals to truth values.

Some quantifier rules refer to eigenvariables restricted to occur only in certain formulae. Isabelle-86 represents such restrictions literally, using a directed graph. The usual data structure for a formula is already a graph; the additional arcs connect the variable $x$ to the formulae where it may not appear, say $Q$ and $R$. A variable instantiation introducing an occurrence of $x$ in $Q$ or $R$ would be detected as a cycle. Logicians may recognize that in this approach, eigenvariables are essentially Henkin constants.

The new meta-logic requires no special mechanism for eigenvariables (Paulson, 1989). The quantifier $\bigwedge$ can express the generality intended by the syntactic restrictions. The premise of $\forall$ introduction is general: that $F(x)$ is true for arbitrary $x$. The corresponding axiom expresses that $\forall$ is an infinitary conjunction, as discussed in Section 6.1.

$$\bigwedge F\,.\,(\bigwedge x\,.\,[\![F(x)]\!]) \Longrightarrow [\![\forall x.F(x)]\!]$$

$$\bigwedge F y\,.\,[\![\forall x.F(x)]\!] \Longrightarrow [\![F(y)]\!]$$

Together these axioms define $[\![\forall x.F(x)]\!]$ as equivalent to $\bigwedge x\,.\,[\![F(x)]\!]$.

The $\exists$ introduction rule is straightforward. But $\exists$ elimination is the most complicated rule of all, having a general premise that discharges an assumption. Let us verify the axiom of $\exists$ elimination using the semantics. If $\exists x.F(x)$ is true, then $F(x)$ is true for some value of $x$. Since $F(x)$ implies $Q$ for all $x$, indeed $Q$ is true.

$$\bigwedge F y\,.\,[\![F(y)]\!] \Longrightarrow [\![\exists x.F(x)]\!]$$

$$\bigwedge F Q\,.\,[\![\exists x.F(x)]\!] \Longrightarrow (\bigwedge x\,.\,[\![F(x)]\!] \Longrightarrow [\![Q]\!]) \Longrightarrow [\![Q]\!]$$

# 7 Soundness and completeness

Is the Isabelle representation of classical logic correct? Each axiom is sound with respect to the truth-table semantics of the logical constants, but we can do better. There is a syntactic, rule-by-rule translation between meta-proofs and object-proofs (indicated by triangles below):

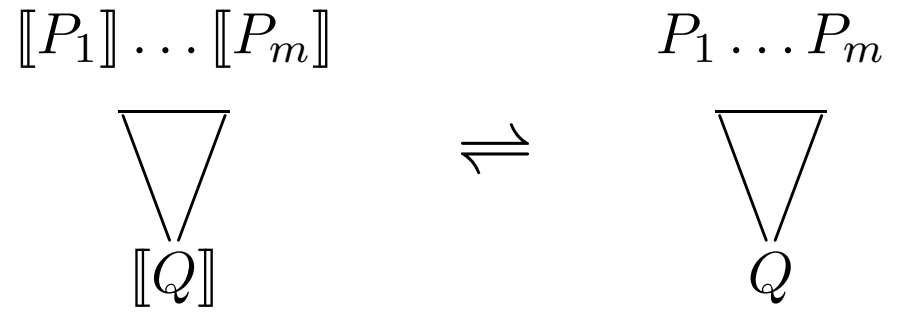

The representation is *sound* if for every meta-proof there is a corresponding object-proof, and *complete* if for every object-proof there is a corresponding meta-proof.

Completeness is easy to demonstrate. We translate object-proofs to meta-proofs by induction on the size of the object-proof. Three cases are shown.

If the last inference is $\vee$ introduction, instantiate the corresponding axiom to the formulae $p$ and $q$. Then use the induction hypothesis to prove $[\![p]\!]$.

$$\dfrac{\dfrac{\dfrac{\bigwedge PQ\,.\,[\![P]\!] \Longrightarrow [\![P \vee Q]\!]}{\bigwedge Q\,.\,[\![p]\!] \Longrightarrow [\![p \vee Q]\!]}}{[\![p]\!] \Longrightarrow [\![p \vee q]\!]} \qquad \begin{array}{c}\bigtriangledown \\ [\![p]\!]\end{array}}{[\![p \vee q]\!]}$$

If the last inference is $\supset$ introduction, then the induction hypothesis yields a proof of $[\![q]\!]$ from the assumption $[\![p]\!]$. Discharging this assumption at the meta-level proves $[\![p]\!] \Longrightarrow [\![q]\!]$.

$$\dfrac{\dfrac{\dfrac{\bigwedge PQ\,.\,([\![P]\!] \Longrightarrow [\![Q]\!]) \Longrightarrow [\![P \supset Q]\!]}{\bigwedge Q\,.\,([\![p]\!] \Longrightarrow [\![Q]\!]) \Longrightarrow [\![p \supset Q]\!]}}{([\![p]\!] \Longrightarrow [\![q]\!]) \Longrightarrow [\![p \supset q]\!]} \qquad \dfrac{\begin{array}{c}\big[\,[\![p]\!]\,\big] \\ \bigtriangledown \\ [\![q]\!]\end{array}}{[\![p]\!] \Longrightarrow [\![q]\!]}}{[\![p \supset q]\!]}$$

For $\forall$ introduction, the scope of the quantifier $(F)$ is instantiated to some abstraction $\lambda x.p[x]$. Then the induction hypothesis proves $[\![\,p[x]\,]\!]$, since $x$ is not free in the assumptions, and so the meta-rule for $\bigwedge$ proves $\bigwedge x\,.\,[\![p]\!]$.

$$\dfrac{\dfrac{\bigwedge F\,.\,(\bigwedge x\,.\,[\![F(x)]\!]) \Longrightarrow [\![\forall x.F(x)]\!]}{(\bigwedge x\,.\,[\![\,p[x]\,]\!]) \Longrightarrow [\![\forall x.p[x]]\!]} \qquad \dfrac{\begin{array}{c}\bigtriangledown \\ [\![\,p[x]\,]\!]\end{array}}{\bigwedge x\,.\,[\![\,p[x]\,]\!]}}{[\![\forall x.p[x]]\!]}$$

Observe that the object and meta-logics share the mechanisms for assumption discharge and quantifier provisos.

Soundness holds because every occurrence of an axiom in a meta-proof must take the form illustrated in the three cases above. By a standard result of proof theory (Prawitz, 1965), every meta-proof can be put into *extended normal form*, consisting of a branch rising to the left and terminating in an axiom. Subproofs have the same structure. A recursive process translates the meta-proof to a corresponding object-proof.

My soundness argument is presented for intuitionistic first-order logic (Paulson, 1989). It appears valid for all similar logics, just the ones that are easy to formalize in Isabelle. Problematical logics (for example, modal) require separate demonstrations of soundness.

## 8 The Edinburgh Logical Framework

Inevitably we must compare Isabelle with the Edinburgh Logical Framework, or LF (Harper et al., 1987). The two meta-logics have much in common, but Isabelle deals only with provability, while the LF formalizes the object-proofs themselves.

## 8.1 Propositions as types

The LF is closely related to a formal system of Per Martin-Löf that is based on the idea of *propositions as types.* This makes use of the general product type. If $A$ is a type and $B[x]$ is a type for all $x \in A$ then the type $\prod_{x\in A} B[x]$ is the product of $B[x]$ over $A$. Its elements are all functions $f$ such that if $a \in A$ then $f(a) \in B[a]$.

Here are the introduction and elimination rules for $\Pi$:

$$\frac{\begin{array}{c}[x \in A] \\ b[x] \in B[x]\end{array}}{\lambda_{x\in A} b[x] \in \prod_{x\in A} B[x]} * \qquad \frac{f \in \prod_{x\in A} B[x] \qquad a \in A}{f(a) \in B[a]}$$

* Proviso: the variable $x$ must not be free in the assumptions save $x \in A$.

The introduction rule says, if $b[x] \in B[x]$ for all $x \in A$ then the function $\lambda_{x\in A} b[x]$ is an element of $\prod_{x\in A} B[x]$. It is sometimes called a *dependent* function since the type of the result may depend on the value of the argument. Strictly speaking, the introduction rule also requires that $A$ is a type, for with indexed families of types it may not be obvious whether a type is well-formed.

If the type $B$ does not depend on $x$ then $\prod_{x\in A} B[x]$ is abbreviated $A \to B$, the type of ordinary functions from $A$ to $B$. In this case the rules simplify to

$$\frac{\begin{array}{c}[x \in A] \\ b[x] \in B\end{array}}{\lambda_{x\in A} b[x] \in A \to B} * \qquad \frac{f \in A \to B \qquad a \in A}{f(a) \in B}$$

The function types $A \to B$ generate the simple types of Isabelle. We furthermore can interpret Isabelle's meta-logic using 'propositions as types'. Each proposition $\phi$ is interpreted by a set of proof terms for $\phi$, following the intuitionistic reading of the logical constants. (See Nordström and Smith (1984) for a discussion.) Here only implication ($\Longrightarrow$) and universal quantification ($\bigwedge$) must be considered.

A proof of $\phi \Longrightarrow \psi$ is a function that maps proofs of $\phi$ to proofs of $\psi$. If $A$ is the type of $\phi$ proofs and $B$ is the type of $\psi$ proofs then $A \to B$ is the type of $\phi \Longrightarrow \psi$ proofs.

A proof of $\bigwedge x\,.\,\psi[x]$, where $x \in A$, is a function that maps each $a \in A$ to a proof of $\psi[a]$. If $B[x]$ is the type of $\psi[x]$ proofs for $x \in A$ then $\prod_{x\in A} B[x]$ is the type of $\bigwedge x\,.\,\psi[x]$ proofs.

This interpretation, based on semantic ideas, is reflected in the syntax of the rules. Compare them with those of Section 5. Ignoring the elements of the types, which represent proof terms, the $\to$ rules resemble the $\Longrightarrow$ rules. If we treat $A$ as an ordinary type, the $\Pi$ rules also resemble the $\bigwedge$ rules. The type constraints for $x$ are implicit in the $\bigwedge$ rules, but explicit in the $\Pi$ rules.

## 8.2 First-order logic in the LF

We now formalize first-order logic in the LF for comparison with Isabelle's approach.

We introduce types *term* and *form* and the logical constants, as before. But we do not have true $\in$ *form* $\to$ *prop*. There is no type of propositions since here propositions themselves are types. Instead, introduce a type-valued function 'true' such that true($P$) is a type for all $P \in$ *form*. Let us continue to abbreviate true($P$) as $[\![P]\!]$, remembering this is now a type.

It is time to formalize the rules. Where Isabelle uses axioms to assert that a formula $P$ is true, the LF uses constants to construct elements of type $[\![P]\!]$. Recall the Isabelle axioms for $\land$ introduction, $\supset$ introduction, and $\forall$ introduction:

$$\bigwedge PQ\,.\,[\![P]\!] \Longrightarrow ([\![Q]\!] \Longrightarrow [\![P \land Q]\!])$$

$$\bigwedge PQ\,.\,([\![P]\!] \Longrightarrow [\![Q]\!]) \Longrightarrow [\![P \supset Q]\!]$$

$$\bigwedge F\,.\,(\bigwedge x\,.\,[\![F(x)]\!]) \Longrightarrow [\![\forall x.F(x)]\!]$$

The corresponding constant declarations are

$$\text{conjI} \in \prod_{P \in \mathit{form}} \prod_{Q \in \mathit{form}} [\![P]\!] \to ([\![Q]\!] \to [\![P \land Q]\!])$$

$$\text{implI} \in \prod_{P \in \mathit{form}} \prod_{Q \in \mathit{form}} ([\![P]\!] \to [\![Q]\!]) \to [\![P \supset Q]\!]$$

$$\text{allI} \in \prod_{F \in \mathit{term} \to \mathit{form}} \Big( \prod_{x \in \mathit{term}} [\![F(x)]\!] \Big) \to [\![\forall x.F(x)]\!]$$

Now let us look at proof construction.

There is a constant for every object-rule, so each derivation produces a term formalizing the object-proof. If proofp is a proof of $p$ and proofq is a proof of $q$ then conjI yields a proof of $p \land q$. Note that conjI must be applied to $p$ and $q$ themselves, as well as their proofs. Applying $\Pi$ elimination four times derives

$$\frac{p \in \mathit{form} \qquad q \in \mathit{form} \qquad \text{proofp} \in [\![p]\!] \qquad \text{proofq} \in [\![q]\!]}{\text{conjI}(p)(q)(\text{proofp})(\text{proofq}) \in [\![p \land q]\!]}$$

The $\supset$ introduction rule discharges an assumption. If proofr$[z]$ is a proof of $r$, where $z$ ranges over proofs of $p$, we can construct a function from $[\![p]\!]$ to $[\![r]\!]$:

$$\frac{\begin{array}{c}\Big[z \in [\![p]\!]\Big] \\ \text{proofr}[z] \in [\![r]\!]\end{array}}{\lambda_{z \in [\![p]\!]}\text{proofr}[z] \in [\![p]\!] \to [\![r]\!]}$$

Applying impI to $p$, $r$, and this function yields a proof of $p \supset r$:

$$\frac{p \in \mathit{form} \qquad r \in \mathit{form} \qquad \begin{array}{c}\Big[z \in [\![p]\!]\Big] \\ \text{proofr}[z] \in [\![r]\!]\end{array}}{\text{impI}(p)(r)(\lambda_{z \in [\![p]\!]}\text{proofr}[z]) \in [\![p \supset r]\!]}$$

For an example of quantifiers, consider $\forall$ introduction. Let $h[x]$ be a formula whose proof is proofh$[x]$ for $x \in \mathit{term}$. Apply $\to$ introduction to form the scope of the quantifier and $\Pi$ introduction to form a proof function:

$$\frac{\begin{array}{c}[x \in \mathit{term}] \\ h[x] \in \mathit{form}\end{array}}{\lambda_{x \in \mathit{term}} h[x] \in \mathit{term} \to \mathit{form}} \qquad \frac{\begin{array}{c}[x \in \mathit{term}] \\ \text{proofh}[x] \in [\![\, h[x] \,]\!]\end{array}}{\lambda_{x \in \mathit{term}}\text{proofh}[z] \in \prod_{x \in \mathit{term}} [\![\, h[x] \,]\!]}$$

Combining these meta-proofs yields

$$\text{allI}(\lambda_{x\in term} h[x])(\lambda_{x\in term}\text{proofh}[z]) \in [\![\forall x \,.\, h[x]]\!]$$

We now consider a complete, if trivial, proof of the tautology $P \supset (P \wedge P)$:

$$\frac{\dfrac{[P] \qquad [P]}{P \wedge P}}{P \supset (P \wedge P)}$$

The proof term is

$$\text{impI}(P)(P \wedge P)\Big(\lambda_{z\in[\![P]\!]}\text{conjI}(P)(P)(z)(z)\Big) \;\in\; [\![P \supset (P \wedge P)]\!]$$

Note that it contains 6 copies of $P$. Proof objects contain an enormous amount of redundancy: just like ordinary 'proof trees'.

### 8.3 Observations

Should we theorize first and code later, or vice versa? Isabelle is a computer program, now several years old. It was first based on elementary ideas. These have now been more formally developed. The LF is a theoretical system for which implementations are emerging.

The main difference between Isabelle's meta-logic and the LF are the proof terms. Because of storage limitations, the first practical applications of the LF will probably be proof editors and checkers, where only small proofs are involved. Isabelle is intended for large, semi-automatic proofs. It constructs proof objects only when, as in Constructive Type Theory, they are explicitly present in the logic.

The essence of Martin-Löf's approach is contained in the two Π rules. The Edinburgh LF appears to be a much larger system: Harper et al. (1987) present 17 rules. The extra rules formalize mundane aspects of the representation of a logic. For example, recall that 'true' is a type-valued function. To express such things, the LF has supertypes, called *Kinds*. The Kind *Type* has all types as elements, while the Kind *form* → *Type* contains functions from type *form* to types. In the LF, we would declare true ∈ *form* → *Type*. Other LF rules construct *signatures*: sequences of constant declarations that define a logic.

Judging by the literature, these extended type theories do not seem to have useful, intuitive models. Martin-Löf uses an informal semantics. The LF researchers abandon the semantical viewpoint altogether, regarding the formulation of a logic as a purely syntactic coding.

Isabelle and the LF appear to deal with the same class of logics. An Isabelle representation (in higher-order logic) can be translated into the LF as sketched above, while Felty and Miller (1988) describe the reverse translation. Avron et al. (1987) have formalized a variety of logics. An LF representation is often more elegant, while Isabelle benefits from the large body of knowledge about higher-order logic.

## 9 Automated Proof in Isabelle

Isabelle supports automated proof through tactics and tacticals which — although implemented in a completely new way — resemble those of Edinburgh LCF and its descendants.

To compare Isabelle's representation of rules with LCF's, consider PROLOG. PROLOG has no functions, only relations; many relations are truly bidirectional, having no distinction between input and output. Similarly Isabelle rules have no inputs or outputs; they describe a relationship between premises and conclusion that must hold in a valid proof. A rule, with its premises and conclusion, has the same basic structure as a Horn clause in PROLOG.

There are different styles of constructing the proof tree: *forwards proof* works from the leaves downwards; *backwards proof* works from the root upwards. The LCF view of rules — as functions from premises to conclusion — is biased towards forwards proof. By adopting a bidirectional view of rules we get a simpler treatment of backwards proof.

Isabelle represents object-rules with assertions of the general form

$$[\phi_1, \ldots, \phi_m] \Longrightarrow \phi$$

Observe that this assertion has two valid readings:

- the *rule* with premises $\phi_1, \ldots, \phi_m$ and conclusion $\phi$
- the *proof state* with subgoals $\phi_1, \ldots, \phi_m$ and final goal $\phi$

The latter reading is that the subgoals $\phi_1, \ldots, \phi_m$ are sufficient to achieve the final goal $\phi$. The initial state is typically the trivial implication $[\![P]\!] \Longrightarrow [\![P]\!]$: here the goal is to prove the formula $P$, and the one subgoal is the same. If we eventually reduce the proof state to $[\![P]\!]$, then the backwards proof is successful — and the desired theorem is the proof state itself. The state of most theorem provers is hidden in the data structures, but Isabelle's proof state is itself subject to the reasoning process.

Rules are essentially Horn clauses. We build proofs by combining two rules, and this is simply Horn clause resolution. Let $[\psi_1, \ldots, \psi_m] \Longrightarrow \psi$ and $[\phi_1, \ldots, \phi_n] \Longrightarrow \phi$ be Horn clauses. Let $s$ be any (higher-order) unifier of $\psi$ and $\phi_i$. Writing $Xs$ for the application of substitution $s$ to the expression $X$, resolution is the following meta-inference:

$$\frac{[\psi_1, \ldots, \psi_m] \Longrightarrow \psi \qquad [\phi_1, \ldots, \phi_n] \Longrightarrow \phi}{[\phi_1, \ldots, \phi_{i-1}, \psi_1, \ldots, \psi_m, \phi_{i+1}, \ldots, \phi_n]s \Longrightarrow \phi s} \qquad (\psi s = \phi_i s)$$

Typically $[\psi_1, \ldots, \psi_m] \Longrightarrow \psi$ is an object-rule, while $[\phi_1, \ldots, \phi_n] \Longrightarrow \phi$ is the proof state. The result is the new proof state with the $i$th subgoal replaced by $\psi_1 s, \ldots, \psi_m s$. Substitution $s$ updates variables in the object-rule, producing the correct instance to apply. It also may update variables in the proof state, solving unknowns in other subgoals and the main goal $\phi$. Since the proof state is an assertion, the role of variables is easy to understand and the effect of an update is immediately visible.

While LCF systems have a tactic for each rule, resolution is a uniform method of inference. The standard tactic `resolve_tac` applies a list of rules to a subgoal. LCF *tacticals* provide sequencing, choice, repetition of tactics. Similar control structures are easily obtained for Isabelle tactics, even though they work on different principles. Most of these tactics and tacticals are independent of the object-logic, which means they are available to all.

The Isabelle notion of tactic allows for an infinite sequence of results, for there may be infinitely many higher-order unifiers in resolution. Multiple results are a useful generalization that could be applied in LCF. They makes it easy to write tacticals that repeatedly apply some tactic and produce a search tree. One standard tactical uses a depth-first strategy, another best-first. The tacticals return an infinite sequence of success nodes. Infinite lazy lists are implemented by the standard closure trick (Paulson, 1987, page 188).

The joining of rules should form a proof, but Isabelle discards the internal structure. Thus the joining of rules derives a new rule. Derived rules can be used exactly like primitive rules — and with the same efficiency. They play an important role in many logics, particularly in set theory, where an entire sequent calculus is derived from the axioms.

Let us return to the example, classical logic. We cannot stop with formalizing the rules and expect the hapless user to be able to prove theorems. Natural deduction is notoriously awkward for classical logic; for starters, try proving $P \vee \neg P$.

Classical proof is much easier with the help of derived rules such as the following (Paulson, 1987, pages 46–49):

$$\frac{\begin{matrix}[\neg Q] \\ P\end{matrix}}{P \vee Q} \qquad \frac{P \supset Q \quad \begin{matrix}[\neg P] \\ R\end{matrix} \quad \begin{matrix}[Q] \\ R\end{matrix}}{R} \qquad \frac{\neg P \quad \begin{matrix}[\neg Q] \\ P\end{matrix}}{Q}$$

The first two of these exploit the classical equivalence of $P \supset Q$ and $\neg P \vee Q$. The third rule is the swap rule. It is typically applied to an assumption $\neg P$, where $P$ is a complex formula, creating a subgoal where $P$ can be broken up using introduction rules.

Such derived rules simulate the sequent calculus LK (Takeuti, 1987), where a sequent $P_1, \ldots, P_m \vdash Q_1, \ldots, Q_n$ can have more than one formula on the right. Why not use LK directly? Indeed I have used LK extensively in Isabelle, but have come to prefer the clumsy treatment shown above. Natural deduction seems to work better because it relies on the standard implication ($\Longrightarrow$). The sequent calculus defines another kind of implication ($\vdash$), and the manipulation of sequents involves a lot of extra work.

Either formalization leads to a complete automatic theorem prover. While *much* slower than standard classical provers, they are fast enough for interactive use, and they are flexible. They are a useful tool for stronger logics like arithmetic and set theory, in proofs by induction.

The original (Isabelle-86) meta-logic is tailor-made for resolution, for rules look like Horn clauses. The switch to higher-order logic has introduced some complications (Paulson, 1989). The axioms that represent object-rules are stripped of their outer quantifiers, leaving implications. A subgoal formula may have a context of assumptions and eigenvariables, meaning it may be enclosed by several occurrences of $\Longrightarrow$ and $\bigwedge$. An object-rule is automatically 'lifted' into this context before it is applied.

These forms of meta-reasoning are derivable within higher-order logic but are hand-coded in Isabelle for speed. The translation of object-proofs into higher-order logic makes them 3–8 times bigger; resolution takes the expansion factor back down to one.

Felty and Miller (1988) formalize logics and theorem provers in $\lambda$Prolog, a version of Prolog that uses higher-order unification. The logical basis of $\lambda$Prolog is extremely close to Isabelle's, and rules are represented as $\lambda$Prolog clauses. The main difference is that tactics are also written in $\lambda$Prolog, not ML.

## 10 Some Isabelle object logics

Now it is time for a close look at some implemented Isabelle logics: intuitionistic logic, Constructive Type Theory, a classical sequent calculus, and set theory. A recent imple-mentation of higher-order logic will be described in a future paper.

The concrete syntax of a logic is specified by writing parsing and printing functions in ML. Here is the computer syntax of the meta-logic symbols.

| | | |
|---|---|---|
| $\bigwedge xy.$ | is | `!(x,y)` |
| $\Longrightarrow$ | is | `==>` |
| $\equiv$ | is | `==` |

Constant and type declarations are omitted. When reading the rules, assume that any one-letter identifier is a variable. For more details, see the User's Manual (Paulson, 1988).

## 10.1 Intuitionistic logic with natural deduction

The theory `NJ` implements intuitionistic first-order logic (Prawitz, 1965). Natural deduction involves a combination of forwards and backwards reasoning, particularly with the elimination rules for $\wedge$, $\supset$, and $\forall$. Alternative forms of these are derived. This yields a logic similar to Gentzen's LJ, a cut-free sequent calculus suited to automatic proof (Takeuti, 1987).

Here are the rules for conjunction, disjunction, and implication:

```
[| P |] ==> [| Q |] ==> [| P&Q |]
[| P&Q |] ==> [| P |]               [| P&Q |] ==> [| Q |]

[| P |] ==> [| P|Q |]               [| Q |] ==> [| P|Q |]

[| P|Q |] ==> ([| P |] ==> [| R |]) ==>
              ([| Q |] ==> [| R |]) ==> [| R |]

([| P |] ==> [| Q |]) ==> [| P-->Q |]
[| P-->Q |] ==> [| P |] ==> [| Q |]
```

Here is how we define $P \leftrightarrow Q$ in terms of conjunction and implication. Abstract rules for $\leftrightarrow$ can then be derived and used as though they were primitive.

```
P<->Q == (P-->Q) & (Q-->P)
```

Here are the quantifier rules.

```
(!(y) [| P(y) |]) ==> [| ALL x.P(x) |]
[| ALL x.P(x) |] ==> [| P(a) |]

[| P(a) |] ==> [| EXISTS x.P(x) |]
[| EXISTS x.P(x) |] ==> (!(y)[| P(y) |] ==> [| R |]) ==> [| R |]
```

This logic is distributed with about 75 sample theorems. Most, including the following, are proved by an automatic tactic.

```
[| (~ ~ P) & ~ ~ (P --> Q) --> (~ ~ Q) |]

[| (EXISTS x. EXISTS y. P(x) & Q(x,y))
       <-> (EXISTS x. P(x) & EXISTS y. Q(x,y)) |]

[| (EXISTS y. ALL x. P(x) --> Q(x,y))
         --> ALL x. P(x) --> EXISTS y. Q(x,y) |]
```

## 10.2 Constructive Type Theory

Isabelle formulates an extensional version of Martin-Löf Type Theory using natural deduction. A typical judgement is expressed using $\bigwedge$ and $\Longrightarrow$:

$$\bigwedge x_1 . \llbracket x_1 \in A_1 \rrbracket \Longrightarrow \bigwedge x_2 . \llbracket x_2 \in A_2(x_1) \rrbracket \Longrightarrow \cdots \bigwedge x_n . \llbracket x_n \in A_n(x_1, \ldots, x_{n-1}) \rrbracket \Longrightarrow \llbracket a(x_1, \ldots, x_n) \in A(x_1, \ldots, x_n) \rrbracket$$

Assumptions can use all the judgement forms, not just $x \in A$, and can even express that $B$ is a family of types over $A$:

$$\bigwedge x . \llbracket x \in A \rrbracket \Longrightarrow \llbracket B(x) \text{ type} \rrbracket$$

Here are some rules for the sum of two types, $A + B$, namely a formation rule, an introduction rule, an elimination rule, and an equality (computation) rule.

```
[| A type |] ==> [| B type |] ==> [| A+B type |]
[| a: A |] ==> [| B type |] ==> [| inl(a): A+B |]

[| p: A+B |] ==> (!(x)[| x: A |] ==> [| c(x): C(inl(x)) |]) ==>
                 (!(y)[| y: B |] ==> [| d(y): C(inr(y)) |]) ==>
                 [| when(p,c,d): C(p) |]

[| a: A |] ==> (!(x)[| x: A |] ==> [| c(x): C(inl(x)) |]) ==>
               (!(y)[| y: B |] ==> [| d(y): C(inr(y)) |]) ==>
               [| when(inl(a),c,d) = c(a): C(inl(a)) |]
```

Here are some of the rules for the general product, $\prod_{x \in A} B(x)$.

```
[| A type |] ==> (!(w)[| w: A |] ==> [| B(w) type |]) ==>
                 [| Prod(A,B) type |]

[| A type |] ==> (!(w)[| w: A |] ==> [| b(w): B(w) |]) ==>
                 [| lambda(b): Prod(A,B) |]

[| p: Prod(A,B) |] ==> [| a: A |] ==> [| p ` a: B(a) |]

[| a: A |] ==> (!(w)[| w: A |] ==> [| b(w): B(w) |]) ==>
               [| lambda(b) ` a = b(a): B(a) |]
```

Constructive Type Theory is distributed with about 100 sample theorems. Tactics based on those of `NJ` can automatically prove many logical statements expressed using propositions as types. In the two examples below the goal contains the variable `?a` standing for some proof object. Isabelle instantiates this variable.

The first example is the derivation of a currying functional. The argument of the functional is a function that maps $z : \Sigma(A, B)$ to $C(z)$; the resulting function maps $x \in A$ and $y \in B(x)$ to $C(\langle x, y \rangle)$. Here is the initial goal; $B$ is a family over $A$ while $C$ is a family over $\Sigma(A, B)$.

```
[| A type |] ==>
(!(x)[| x:A |] ==> [| B(x) type |]) ==>
(!(z)[| z: (SUM x:A . B(x)) |] ==> [| C(z) type |]) ==>
[| ?a: (PROD z: (SUM x:A . B(x)) . C(z))
                  --> (PROD x:A . PROD y:B(x) . C(<x,y>)) |]
```

This goal is proved in one step by an automatic tactic, replacing `?a` by the currying functional

```
lam ka. lam kb. lam kc. ka ‘ <kb,kc>
```

The second example is a strong choice principle (Martin-Löf, 1984, page 50). The proof requires a complicated series of commands. The initial goal is

```
[| A type |] ==>
(!(x)[| x:A |] ==> [| B(x) type |]) ==>
(!(x,y)[| x:A |] ==> [| y:B(x) |] ==> [| C(x,y) type |]) ==>
[| ?a: (PROD x:A. SUM y:B(x). C(x,y))
       --> (SUM f: (PROD x:A. B(x)). PROD x:A. C(x, f‘x))  |]
```

By the end of the proof, `?a` has become

```
lam ka. <lam u. fst(ka‘u), lam kb. snd(ka‘kb)>
```

Another collection of examples develops elementary arithmetic. Theorems include $m+n = n+m$, $m\times n = n\times m$, $(m+n)\times k = m\times k+n\times k$, and $(m\times n)\times k = m\times(n\times k)$, culminating in

$$m \bmod n + (m/n) \times n = m$$

## 10.3 Classical first-order logic

The theory `LK` implements classical first-order logic using the sequent calculus. Assertions have the form $\Gamma \vdash \Delta$, where $\Gamma$ and $\Delta$ are lists of formulae. A dollar sign prefix (`$`) indicates a sequence variable.

Here are some structural rules: basic sequents, thinning, and cut.

```
[| $H, P, $G |- $E, P, $F |]
[| $H |- $E, $F |] ==> [| $H |- $E, P, $F |]
[| $H |- $E, P |] ==> [| $H, P |- $E |] ==> [| $H |- $E |]
```

These are the rules for conjunction and negation:

```
[| $H |- $E, P, $F |] ==> [| $H |- $E, Q, $F |] ==>
                          [| $H |- $E, P&Q, $F |]
[| $H, P, Q, $G |- $E |] ==> [| $H, P&Q, $G |- $E |]

[| $H, P |- $E, $F |] ==> [| $H |- $E, ~P, $F |]
[| $H, $G |- $E, P |] ==> [| $H, ~P, $G |- $E |]
```

These are the rules for the universal quantifier:

```
(!(x)[| $H |- $E, P(x), $F |]) ==> [| $H |- $E, Forall(P), $F |]

[| $H, P(a), $G, Forall(P) |- $E |] ==>
            [| $H, Forall(P), $G |- $E |]
```

Around 75 examples are distributed. Most, like these, are proved automatically.

```
[| H |- (ALL x. ALL y. EXISTS z. ALL w.(P(x)&Q(y)-->R(z)&S(w)))
       --> (EXISTS x. EXISTS y. P(x) & Q(y)) --> EXISTS z.R(z) |]

[|  EXISTS x. P(x) & ~Q(x),
    ALL x. P(x) --> R(x),
    ALL x. M(x) & L(x) --> P(x),
    (EXISTS x. R(x) & ~ Q(x)) --> (ALL x. L(x) --> ~ R(x))
 |- ALL x. M(x) --> ~L(x) |]
```

## 10.4 Zermelo-Fraenkel set theory

The theory called `set` implements Zermelo-Fraenkel set theory over `LK`.

The following are unusual definitions where one *formula* is defined as equal to another. The extensionality axiom states that $A = B$ is equivalent to $A \subseteq B \wedge B \subseteq A$. The power set axiom states that $A \in \text{Pow}(B)$ is equivalent to $A \subseteq B$. They could instead be expressed using 'if and only if' ($\leftrightarrow$).

```
      A<=B  ==  ALL x. x:A --> x:B
       A=B  ==  A<=B & B<=A
 a: (b::B)  ==  a=b | a:B
 A: Pow(B)  ==  A<=B
```

The constant :: satisfies $a :: B = \{a\} \cup B$, constructing finite sets:

$$\{a, b, c, d\} \quad = \quad a :: (b :: (c :: (d :: \emptyset)))$$

Zermelo-Fraenkel set theory permits limited comprehension. By the separation axiom, the set Collect($A$,$P$) forms the set of all $x \in A$ that satisfy $P(x)$. By the replacement axiom, the set Replace($f$,$A$) forms the set of all $f(x)$ for $x \in A$. There are notations for three kinds of comprehension: separation, replacement, and both together.

| *Isabelle notation* | *expansion* | *standard notation* |
|---|---|---|
| [ $x$ \|\| $x$:$A$, $P(x)$ ] | Collect($A$,$P$) | $\{x \in A \mid P[x]\}$ |
| [ $f(x)$ \|\| $x$:$A$ ] | Replace($f$,$A$) | $\{f[x] \mid x \in A\}$ |
| [ $f(x)$ \|\| $x$:$A$,$P(x)$ ] | Replace($f$,Collect($A$,$P$)) | $\{f[x] \mid x \in A \wedge P[x]\}$ |

Because selection and replacement are axiom schemes, there is no finite axiom system for Zermelo-Fraenkel. Isabelle can formalize axiom schemes using function variables:

```
a: Collect(A,P)  ==  a:A & P(a)
c: Replace(f,B)  == EXISTS a. a:B & c=f(a)
```

Besides 2-place union and intersection ($A \cup B$ and $A \cap B$) we have 'big union' and 'big intersection' operators ($\bigcup C$ and $\bigcap C$). These operate on a set of sets; $\bigcup C$ can also be written $\bigcup_{A \in C} A$. Of these operators only 'big union' is primitive. Here are the axioms for union and intersection.

```
A: Union(C)  ==  EXISTS B. A:B  &  B:C
     a Un b  ==  Union({a,b})
   Inter(C)  ==  [ x || x: Union(C), ALL y. y:C --> x:y ]
    a Int b  ==  [ x || x:a, x:b ]
```

Since some of the axioms are unnatural, Isabelle provides derived rules for set theory. Here are a pair of rules for comprehension. For example, if $a \in \{x \in A \mid P[x]\}$ then both $a \in A$ and $P(a)$ hold.

```
[| $H |- $E, a:A, $F |] ==> [| $H |- $E, P(a), $F |] ==>
                            [| $H |- $E, a: Collect(A,P), $F |]

[| $H, a: A, P(a), $G |- $E |] ==>
         [| $H, a: Collect(A,P), $G |- $E |]
```

Here are rules for the subset relation. To show $A \subseteq B$, show that if $x \in A$ then $x \in B$ for arbitrary $x$.

```
(!(x)[| $H, x:A |- $E, x:B, $F |]) ==> [| $H |- $E, A <= B, $F |]

[| $H, $G |- $E, c:A |] ==> [| $H, c:B, $G |- $E |] ==>
                            [| $H, A <= B, $G |- $E |]
```

There are about 60 examples. Many are proved automatically, like this one:

```
[| |- C<=A  <->  (A Int B) Un C = A Int (B Un C) |]
```

Proofs about 'big intersection' tend to be complicated because $\bigcap$ is ill-behaved on the empty set. Two interesting examples are

```
[|  |- Inter(A Un B) = Inter(A) Int Inter(B), A<=0, B<=0 |]

[| ~(C<=0) |- Inter([ A(x) Int B(x) || x:C ]) =
       Inter([ A(x) || x:C ])  Int  Inter([ B(x) || x:C ]) |]
```

In traditional notation these are

$$A \neq \emptyset \wedge B \neq \emptyset \supset \bigcap(A \cup B) = (\bigcap A) \cap (\bigcap B)$$

$$C \neq \emptyset \supset \bigcap_{x \in C} (A(x) \cap B(x)) = (\bigcap_{x \in C} A(x)) \cap (\bigcap_{x \in C} B(x))$$

These proofs require complicated tactics.

Another large example justifies the standard definition of pairing:

$$\langle a, b \rangle \quad = \quad \{\{a\}, \{a, b\}\}$$

It proves that $\langle a, b \rangle = \langle c, d \rangle$ implies $a = c$ and $b = d$. If you think this looks easy, try proving it yourself. The Isabelle proof involves a long series of lemmas.

# 11 Conclusions

It is too early to tell whether Isabelle can compete with specialized systems like Nuprl and HOL in large proofs. However, people are experimenting with Isabelle at a number of sites. There are some results to report.

Philippe Noël, working in Zermelo-Fraenkel set theory, has developed the foundations of fixedpoint theory, including some elementary domain constructions. These proofs contain an enormous amount of detail, for while set theory is universal, its representation of functions and relations is cumbersome. Perhaps set theory is not a practical formal system; at the very least, it requires a more sophisticated proof procedure.

Tobias Nipkow (1989), using logic programming ideas, has implemented a family of rewriting tactics in Isabelle. They resemble tactics provided with Constructive Type Theory but are more powerful. For example, they permit matching under commutative or associative operators. As usual, these tactics are much slower than the algorithms of specialized rewrite rule theorem provers. Their advantages are correctness, simplicity, and applicability in a range of logics. Nipkow (1989) illustrates them through proofs about arithmetic.

The computer is a wonderful experimental tool. Isabelle was not designed; it evolved. Not everyone likes this idea. Specification experts rightly abhor trial-and-error programming. They suggest that no one should write a program without first writing a complete formal specification. But university departments are not software houses. Programs like Isabelle are not products: when they have served their purpose, they are discarded.

Isabelle's user interface is no advance over LCF's, which is widely condemned as 'user-unfriendly': hard to use, bewildering to beginners. Hence the interest in *proof editors*, where a proof can be constructed and modified rule-by-rule using windows, mouse, and menus. But Edinburgh LCF was invented because real proofs require millions of inferences. Sophisticated tools — rules, tactics and tacticals, the language ML, the logics themselves — are hard to learn, yet they are essential. We may demand a mouse, but we need better education and training.

**Acknowledgements.** Isabelle clearly depends on the work of Per Martin-Löf and Robin Milner. Other sources of ideas include Thierry Coquand, Michael Gordon, Gérard Huet, Dale Miller, and Lincoln Wallen. David Matthews's ML compiler has been indispensable. Users of Isabelle, including Philippe de Groote, Tobias Nipkow, and Philippe Noël, made many contributions. Martin Coen, Avra Cohn, and Jeff Joyce commented on the text. The SERC provided a decisive 'efficiency improvement' in the form of workstations (grant GR/E 0355.7). I am grateful to the above people and to anyone mistakenly omitted.

## References

Peter B. Andrews, Dale A. Miller, Eve L. Cohen, and Frank Pfenning (1984). Automating higher-order logic. In W. W. Bledsoe and D. W. Loveland, editors, *Automated Theorem Proving: After 25 Years*, pages 169–192, American Mathematical Society.

Peter B. Andrews (1986). *An Introduction to Mathematical Logic and Type Theory: To Truth Through Proof.* Academic Press.

A. Avron, F. A. Honsell, and I. A. Mason (1987). *Using typed lambda calculus to implement formal systems on a machine.* Report ECS-LFCS-87-31. Computer Science Department, University of Edinburgh.

Graham Birtwistle and P. A. Subrahmanyam (1988), editors. *VLSI Specification, Verification and Synthesis.* Kluwer Academic Publishers.

Alonzo Church (1940). A formulation of the simple theory of types. *Journal of Symbolic Logic*, 5:56–68.

R. L. Constable et al. (1986). *Implementing Mathematics with the Nuprl Proof Development System.* Prentice-Hall International.

N. G. de Bruijn (1972). Lambda calculus notation with nameless dummies, a tool for automatic formula manipulation, with application to the Church-Rosser Theorem. *Indagationes Mathematicae*, 34:381–392.

Amy Felty and Dale Miller (1988). Specifying theorem provers in a higher-order logic programming language. In E. Lusk and R. Overbeek, editors, *9th International Conference on Automated Deduction*, pages 61–80, Springer-Verlag. LNCS 310.

Michael J. C. Gordon (1988). HOL: a proof generating system for higher-order logic. In Birtwistle and Subrahmanyam (1988), pages 73–128.

R. Harper, F. Honsell and G. Plotkin (1987). A Framework for Defining Logics. *Symposium on Logic in Computer Science.* IEEE Computer Society Press, 194–204.

G. P. Huet (1975). A unification algorithm for typed λ-calculus. *Theoretical Computer Science*, 1:27–57.

G. P. Huet and B. Lang (1978). Proving and applying program transformations expressed with second-order patterns. *Acta Informatica*, 11:31–55.

Per Martin-Löf (1984). *Intuitionistic type theory.* Bibliopolis.

Tobias Nipkow (1989). Equational reasoning in Isabelle. *Science of Computer Programming*, 12:123–149.

B. Nordström and J. M. Smith (1984). Propositions and specifications of programs in Martin-Löf's type theory. *BIT*, 24:288–301.

L. C. Paulson (1986). Natural deduction as higher-order resolution. *Journal of Logic Programming*, 3:237–258.

L. C. Paulson (1987). *Logic and Computation: Interactive proof with Cambridge LCF.* Cambridge University Press.

L. C. Paulson (1988). *A preliminary user's manual for Isabelle.* Technical Report 133, University of Cambridge Computer Laboratory.

L. C. Paulson (1989). The foundation of a generic theorem prover. *Journal of Automated Reasoning*, 5:363–397.

F. J. Pelletier (1986). Seventy-five problems for testing automatic theorem provers. *Journal of Automated Reasoning*, 2:191–216. Errata, JAR 4 (1988), 236–236.

D. Prawitz (1965). *Natural Deduction: A Proof-theoretical Study.* Almquist and Wiksell.

P. Schroeder-Heister (1984). Generalized rules for quantifiers and the completeness of the intuitionistic operators &, $\vee$, $\supset$, $\bot$, $\forall$, $\exists$. In *Computation and Proof Theory: Logic*

*Colloquium '83*, pages 399–426, Springer-Verlag. Lecture Notes in Mathematics 1104.

Stefan Sokołowski (1987). Soundness of Hoare's logic: an automatic proof using LCF. *ACM Transactions on Programming Languages and Systems*, 9:100–120.

G. Takeuti (1987). *Proof Theory*, 2nd edition. North Holland.

Å. Wikström (1987). *Functional Programming using ML*. Prentice-Hall International.